\documentclass[aps,prl,longbibliography,reprint,superscriptaddress]{revtex4-2}
\usepackage[colorlinks=true,linkcolor=magenta,urlcolor=blue,citecolor=blue]{hyperref}
\usepackage{amsmath,amssymb,amsfonts,graphicx,tabularx,dcolumn,bm,xcolor,times,float,algorithm,algorithmic,tikz}
\usetikzlibrary{quantikz}

\begin{document}
\title{Markovian Quantum Neuroevolution for Machine Learning}
\author{Zhide Lu}
\thanks{These authors contributed equally to this work.}
\affiliation{Center for Quantum Information, IIIS, Tsinghua University, Beijing 100084, People’s Republic of China}

\author{Pei-Xin Shen}
\thanks{These authors contributed equally to this work.}
\affiliation{Center for Quantum Information, IIIS, Tsinghua University, Beijing 100084, People’s Republic of China}

\author{Dong-Ling Deng}
\email{dldeng@tsinghua.edu.cn}
\affiliation{Center for Quantum Information, IIIS, Tsinghua University, Beijing 100084, People’s Republic of China}
\affiliation{Shanghai Qi Zhi Institute, 41th Floor, AI Tower, No. 701 Yunjin Road, Xuhui District, Shanghai 200232, China}
\date{\today}

\begin{abstract}
Neuroevolution, a field that draws inspiration from the evolution of brains in nature, harnesses evolutionary algorithms to construct artificial neural networks. It bears a number of intriguing capabilities that are  typically inaccessible to gradient-based approaches, including optimizing neural-network architectures, hyperparameters, and even learning the training rules. 
In this paper, we introduce a quantum neuroevolution algorithm that autonomously finds near-optimal quantum neural networks for different machine-learning tasks. In particular, we establish a one-to-one mapping between quantum circuits and directed graphs, and reduce the problem of finding the appropriate gate sequences to a task of searching  suitable paths in the corresponding graph as a Markovian process. We benchmark the effectiveness of the introduced algorithm  through concrete examples including classifications of  real-life  images and symmetry-protected topological states. 
Our results showcase the vast potential of neuroevolution algorithms in quantum architecture search, which would boost the exploration towards quantum-learning advantage with noisy intermediate-scale quantum devices.
\end{abstract}

\maketitle

\section{Introduction}
Quantum machine learning studies the interplay between machine learning and quantum physics \cite{DasSarma2019Machine,Biamonte2017Quantum,Dunjko2018Machine,Carleo2019Machine}. On the one hand, machine learning has achieved dramatic success over the past two decades \cite{Lecun2015Deep,Jordan2015Machine} and many problems that were notoriously challenging for artificial intelligence, such as playing the game of Go \cite{Silver2016Mastering,Silver2017Mastering}  or predicting protein structures \cite{senior2020improved},  have been cracked recently.  This gives rise to opportunities for using machine-learning techniques to solve difficult problems in quantum science. Indeed, machine-learning ideas and tools have been invoked in various applications in quantum physics, including representing quantum many-body states \cite{Carleo2017Solving,Gao2017Efficient}, quantum-state tomography \cite{Torlai2018Neuralnetwork, Carrasquilla2019Reconstructing}, nonlocality detection \cite{Deng2018Machine}, topological quantum compiling \cite{Zhang2020Topological}, and learning phases of matter \cite{Zhang2017Quantum,Carrasquilla2017Machine,vanNieuwenburg2017Learning,Wang2016Discovering,Broecker2017Machine,Chng2017Machine,Zhang2017Machine, Wetzel2017Unsupervised,Hu2017Discovering, Zhang2019Machine,Lian2019Machine}, etc. On the other hand, the idea of quantum computing has revolutionized the theories and implementations of computation \cite{Nielsen2010Quantum}. Alternative quantum algorithms may offer unprecedented prospects to enhance, speed up, or innovate machine learning as well \cite{Harrow2009Quantum,Lloyd2014Quantum,Dunjko2016QuantumEnhanced, Amin2018Quantum,Gao2018Quantum,Lloyd2018Quantum,Hu2019Quantum,Schuld2019Quantum}. Without a doubt, the studies of the interplay between machine learning and quantum physics will benefit both fields and the emergent research frontier of quantum machine learning has become  one of today’s most rapidly growing interdisciplinary fields \cite{DasSarma2019Machine,Biamonte2017Quantum,Dunjko2018Machine,Carleo2019Machine}.

\begin{table}
\begin{ruledtabular}
\begin{tabular}{ccc}
    System &
    \parbox[c]{11.5em}{\includegraphics[width=0.2\textwidth]{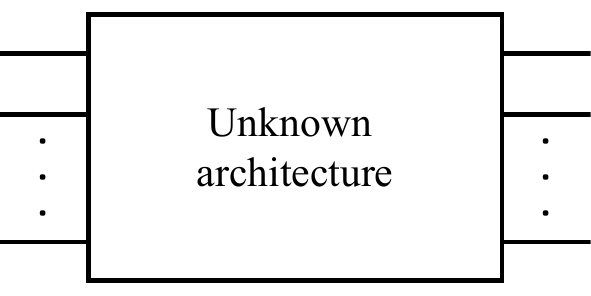}} & 
    \parbox[c]{10em}{\includegraphics[width=0.1\textwidth]{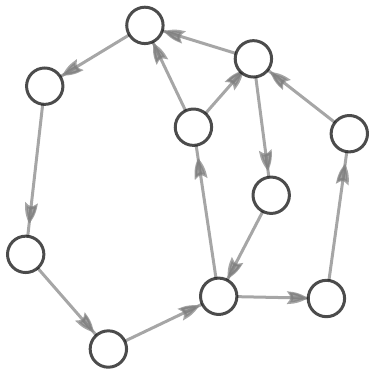}} \\ \hline
    Gate block & 
    \parbox[c]{11.5em}{\includegraphics[width=0.2\textwidth]{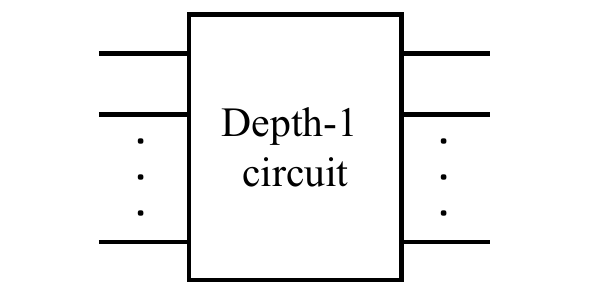}} & 
    \parbox[c]{10em}{\includegraphics[width=0.1\textwidth]{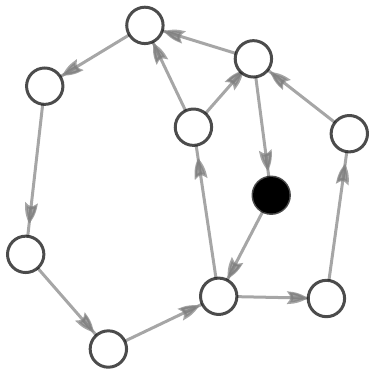}} \\ \hline
    $\begin{array}{c}
        \text{Relation} \\ \text{between blocks}
    \end{array}$ & 
    \parbox[c]{11.5em}{\includegraphics[width=0.2\textwidth]{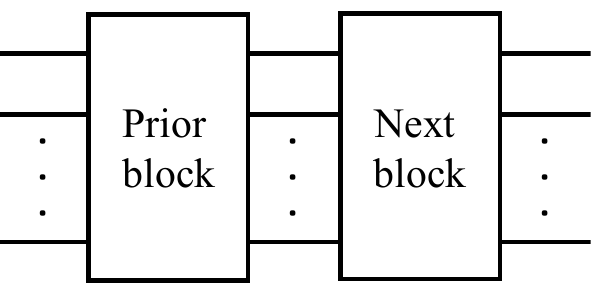}} & $
    \begin{pmatrix}
        0 & 1 & \cdots & 0 \\
        0 & 0 & \cdots & 1 \\
        \vdots & \vdots & \ddots &  \vdots\\
        1 & 0 & \cdots & 0
    \end{pmatrix}$ \\ \hline
    Circuit sequence & 
    \parbox[c]{11.5em}{\includegraphics[width=0.2\textwidth]{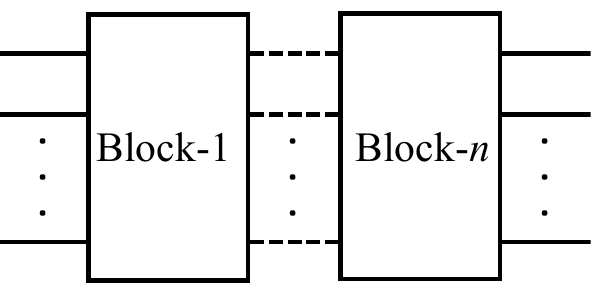}} & 
    \parbox[c]{10em}{\includegraphics[width=0.1\textwidth]{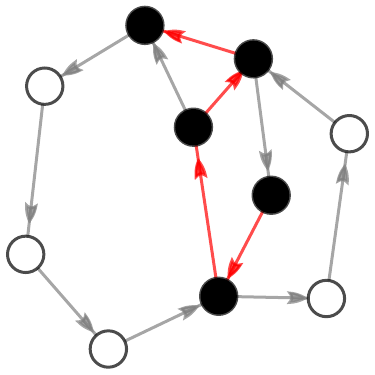}} \\
\end{tabular}
\end{ruledtabular}
\caption{Illustration of the graph-encoding method, based on which the problem of searching the optimal quantum circuits is reduced to a task of finding paths in the corresponding directed graphs. Each node represents one gate-block, and each path (red arrows) represents a sequence of gate-blocks. The nodes that the path passes through are solid, while other nodes are empty.}
\label{tab:comparison betwwen graph and circuit sequences}
\end{table}

An intriguing approach widely studied in quantum machine learning is to exploit the hybrid quantum-classical scheme, where parameterized quantum circuits are optimized with classical methods (such as stochastic gradient descent) to satisfy certain  objective functions.  Notable examples in this category include various quantum classifiers \cite{Schuld2020Circuitcentric,Farhi2018Classification,Schuld2017Implementing,Mitarai2018Quantum,Schuld2019Quantum,Havlicek2019Supervised,Zhu2019Training,Cong2019Quantum,Wan2017Quantum,Grant2018Hierarchical,Du2018Implementable,Uvarov2020Machine,Blank2020Quantum,Rebentrost2014Quantum,Tacchino2019Artificial}, variational quantum eigensolvers \cite{Peruzzo2014Variational,Kokail2019Selfverifying,Liu2019Variational,Wang2019Accelerated}, quantum Born machines \cite{Liu2018Differentiable,Coyle2020Born}, and quantum approximation optimization algorithms  \cite{Farhi2014Quantum,Zhou2020Quantum,Moll2018Quantum}. In this scenario, one typically chooses a variational ansatz circuit with a fixed structure and then optimizes its tunable parameters to tackle the given problem. Yet, different families of parameterized quantum circuits may bear distinct entangling capabilities and representation power, and thus are suitable for different tasks. For a given learning task, how to obtain a well-performing ansatz circuit as short as possible is extremely useful, especially for quantum learning with noisy intermediate-scale quantum (NISQ) devices \cite{Preskill2018Quantum}, where the depth of the quantum circuits would be limited due to undesirable noises carried by such a device. In the classical machine-learning literature, several renowned algorithms have been proposed to search for appropriate neural-network architectures \cite{Real2017LargeScale, Real2019Regularized, Stanley2002Evolving,  Stanley2019Designing,Huang2018GNAS,Zoph2017Neural,Baker2017Designing,Cai2017Efficient,Zoph2018Learning,Liu2019DARTS,Xie2020SNAS,Zela2020Understanding,Liang2020DARTS}, including evolutionary  or genetic algorithms  (such as NeuroEvolution of Augmenting Topologies, NEAT) \cite{Stanley2002Evolving}, greedy algorithms \cite{Huang2018GNAS}, reinforcement learning-based algorithms \cite{Zoph2017Neural,Baker2017Designing,Cai2017Efficient,Zoph2018Learning}, and differentiable architecture search \cite{Liu2019DARTS,Xie2020SNAS,Zela2020Understanding,
Liang2020DARTS}.  Inspired by these algorithms, analogous quantum architecture search algorithms have also been introduced  \cite{Li2017Approximate,Cincio2018Learning,Fosel2018Reinforcement,Rattew2020Domainagnostic,Chivilikhin2020MogVQE,Cincio2020Machine,Ostaszewski2019Quantum,Li2020Quantum,Zhang2020Differentiable,Pirhooshyaran2020Quantum}. Each of these algorithms carries its own pros and cons, and the choice depends on the specific problem.

In this paper, we introduce a quantum neuroevolution algorithm, which we call the Markovian quantum neuroevolution (MQNE) algorithm, to search for optimal ansatz quantum circuits for different machine-learning tasks. We propose a graph-encoding method (see Table \ref{tab:comparison betwwen graph and circuit sequences}), where the nodes of the graph correspond to the elementary gate blocks and the directed edges represent the allowed connection between gate blocks, to injectively map quantum circuits to directed graphs. Consequently, we recast the problem to a task of searching an appropriate directed path of the graph in a Markovian fashion. 
To illustrate the effectiveness of the MQNE algorithm, we apply it to a variety of quantum-learning tasks, including classifications of real-life images (such as handwritten digit images in the MNIST dataset \cite{mnist}, and the Wisconsin Diagnostic Breast Cancer dataset \cite{cancer}) and symmetry-protected topological (SPT) states. We find that our algorithm yields ansatz quantum circuits with notably smaller depths, while maintaining a comparable classification accuracy. 

\begin{table}
\includegraphics[width=0.48\textwidth]{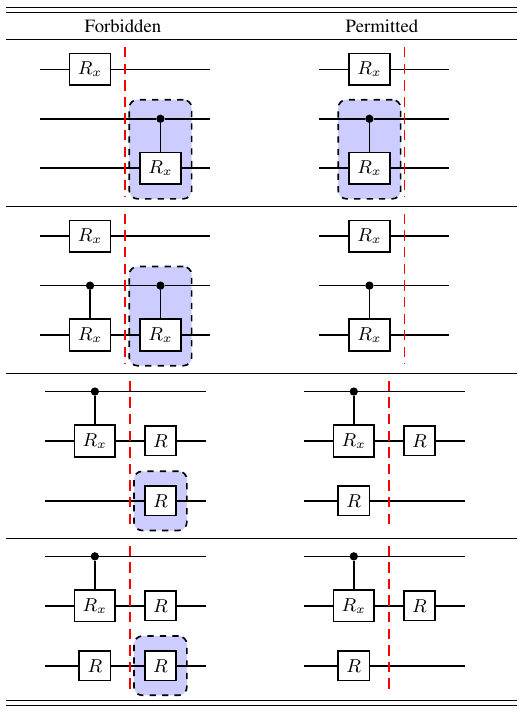}
\caption{ The connection rules for gate blocks (drawn using Quantikz \cite{Kay2019Tutorial}). The left column shows the forbidden connection configurations, which should be replaced by the corresponding permitted ones in the right column. Here, only two-qubit controlled-$R_x$ gate and single-qubit rotation gate $R$ are used in constructing various quantum circuits. }
\label{Table:Rules}
\end{table}

\section{The MQNE algorithm}

In designing classical neural networks, a renowned neuroevolution algorithm is the NEAT algorithm \cite{Stanley2002Evolving}, which exploits concepts (e.g., genome, crossover, speciation, and mutation) from biology to evolve neural-network topologies along with weights. However, straightforward adoption of NEAT in the quantum domain would not work  since quantum neural networks differ substantially from  classical ones. For instance, the quantum neurons (qubits) are connected by multiqubit unitaries rather than weight parameters. As a result, certain techniques, such as explicit fitness sharing and matching up genomes \cite{Stanley2002Evolving}, used in NEAT become invalid or ambiguous in the quantum scenario. Indeed, as shown in the Supplemental Material \footnote{See Supplemental Material at [URL will be inserted by publisher] for details on the graph-encoding method and the MQNE algorithm, and more numerical results to demonstrate the performance of the proposed scheme, which include Refs.~\cite{Bezanson2017Julia,Luo2020Yao}}, the simple genetic algorithm for designing quantum classifiers, which uses crossover and mutation directly, performs poorly in classifying images.  The ineffectiveness of this algorithm is due to the following: i) the encoding of the quantum circuits into bit strings is not a bijection, which increases the search space and slows down the searching process; ii) the performance of the offspring generated from crossover and mutation is not guaranteed to be better than that of their parents, since crossover and mutation of unitaries may result in meaningless structures. 

\begin{figure}
\includegraphics[width=0.48\textwidth]{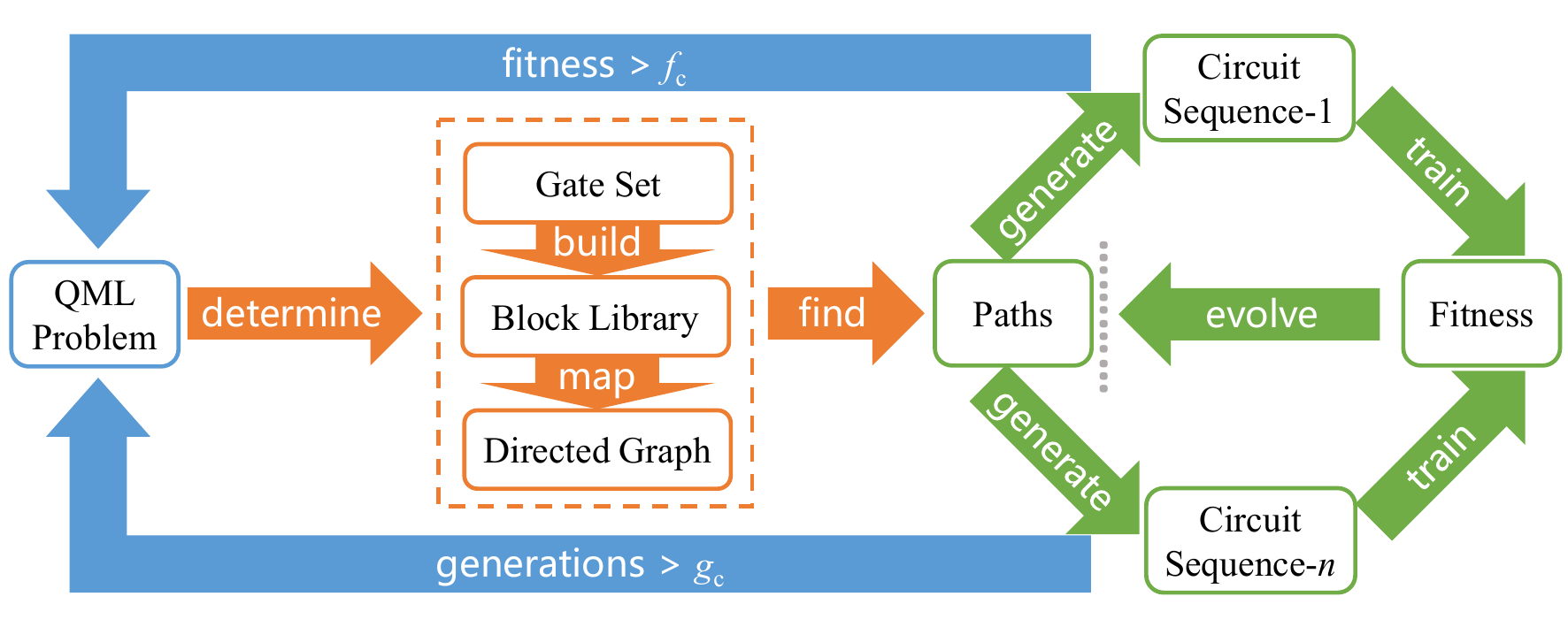}
\caption{Schematic illustration of the MQNE algorithm. For a given quantum machine-learning (QML) problem and experimental setup, we first determine the allowed gate blocks and the corresponding directed graph, and iteratively generate quantum circuits in a Markovian fashion. The algorithm terminates when the highest fitness of the generated circuits becomes larger than a certain threshold value $f_c$ or the number of generations exceeds a given number $g_c$.
\label{fig:algorithm schematic}}
\end{figure}

Our MQNE algorithm overcomes these shortcomings. 
First, we introduce a graph-encoding method, which maps quantum circuits to directed paths in the corresponding graph.  Suppose we need to design a $k$-qubit quantum circuit to solve a given quantum machine-learning problem, and for simplicity we restrict our discussion to the case that the circuits are composed with only single-qubit rotations $R$ and two-qubit controlled-$R_x$ gates. The universal single-qubit rotation gate $R$ is defined as the gate composed of $R_z R_x R_z$, i.e., the Z-X decomposition for a single-qubit rotation, where $R_z$ and $R_x$ denote rotations along $z$ and $x$ axis, respectively \cite{Nielsen2010Quantum}. 
We choose the controlled-$R_x$ gate, rather than the controlled NOT gate typically used in designing quantum neural networks, to guarantee  that the circuits from later generations cover these from earlier generations, so as to ensure improved performance of the offsprings. This can be easily deduced from the fact that the controlled-$R_x$ gate reduces to identity when setting the controlled rotation angle to $0$. To avoid ambiguity and duplication of successive rotations, we invoke some connection rules for arranging gate blocks (a gate block is a depth-1 quantum circuit) in sequential order to form the desired circuits (see Table \ref{Table:Rules}): i) the latter gate-block should not include any gate which can be operated in parallel with the former one; ii) the latter gate block should not include the same gates as the former one on the same qubits. Such connection rules restrict the process of searching the optimal circuit structure in a Markovian fashion. We suppose that the qubits are arranged in a one-dimensional geometry and the controlled-$R_x$ gates act only on adjacent qubits. We use a length-$(k+2\lfloor k/2\rfloor)$ vector to represent a quantum gate block.
The first $2\lfloor k/2\rfloor$ numbers encode controlled-$R_x$ gates in a gate block.
Here, two adjacent nonzero numbers represent a controlled-$R_x$ gate acting on these two qubits labeled by them, and two adjacent $0$ numbers mean that there is no  controlled-$R_x$ gate acting on the remaining qubits. 
The next $k$ numbers encode the single-qubit rotation gates in a gate block, where we use $0$ to denote the absence of rotation for the corresponding qubit \cite{Note1}.

Without further restrictions, it is straightforward to obtain that the number of possible gate-blocks is $f_1(k)=\frac{(1+\sqrt{3})^{k+1}-(1-\sqrt{3})^{k+1}}{2\sqrt{3}}$ \cite{Note1}. These gate-blocks form a gate-block library and we use a directed graph to represent this library. Each node of the graph corresponds to a gate block, and each directed edge represents a legitimate connection of gate blocks according to the connection rules: there is an edge pointing from node $x$ to $y$ if and only if the gate block $y$ is allowed to be put next to gate block $x$. For convenience, we use an adjacency matrix to denote the directed graph as in graph theory \cite{Deo2016Graph}. Noting that a quantum circuit is just a sequence of gate blocks in the corresponding library, hence the task of designing a well-performing quantum circuit is reduced to finding an optimal path in the directed graph. This can be solved with the following procedure: 1) Initialization. we start from a fixed node and uniformly sample $n_1$ paths with length $l$ based on the directed graph, and compute the fitness (classification accuracy) of the corresponding variational quantum circuits. These $n_1$ paths form the first generation.
2) Iteration in a Markovian fashion. From the $i$-th generation, we choose $t_i$ paths with the largest fitness. For each of the selected paths, we uniformly sample $n_{i+1}$ segments of length $l'$, and then add these segments to the end of the path. Here we remark that one segment refers to a sequence of $l'$ gate blocks. Due to the connection rules, not all possible segments can be added at will to the existed paths. Whether a new segment is allowed to be added depends on the last gate block of the given paths, which is similar to a Markovian process. In this way, we obtain the paths of the $(i+1)$-th generation. We then parallel evaluate the fitness of all $(i+1)$-th generation quantum circuits. whose running time is independent of the number of paths at each generation. 
If the fitness of a circuit is larger than a certain given threshold value $f_c$ (or the number of iteration exceeds a given number $g_c$), we terminate the iteration and output the corresponding path and quantum circuit. If none of 
the circuits has a fitness larger than $f_c$, we repeat this step to generate paths and circuits for the next generation.
A schematic illustration for our MQNE algorithm is given in Fig.~\ref{fig:algorithm schematic}, with the pseudocode provided within the Supplemental Material \cite{Note1}.

We note that the number of nodes of the directed graph scales exponentially with the number of qubits involved $f_1(k)= \Theta(2^k)$. For large $k$, the size of the graph might exceed the capacity of any classical computer, rendering our MQNE algorithm infeasible in practice. To reduce the size of the graph, we can impose some further restrictions on building possible gate blocks. For instance, we may require that for each gate block there are at most $c$ (a cutoff constant number) controlled-$R_x$ gates, and the rest qubits all undergo single-qubit rotations.  With these restrictions, the number of possible gate blocks reduces to a polynomial function of $k$ \cite{Note1}. Accordingly, the size of the directed graph is also reduced. However, it is worthwhile to mention that the reduction of the graph may also bring up a problem: we may not be able to find the optimal ansatz circuits since the searching space is reduced too much by the restrictions. 
In the following, we give a couple of concrete examples to benchmark the effectiveness of our MQNE algorithm.

\begin{figure}
\includegraphics[width = 0.48\textwidth]{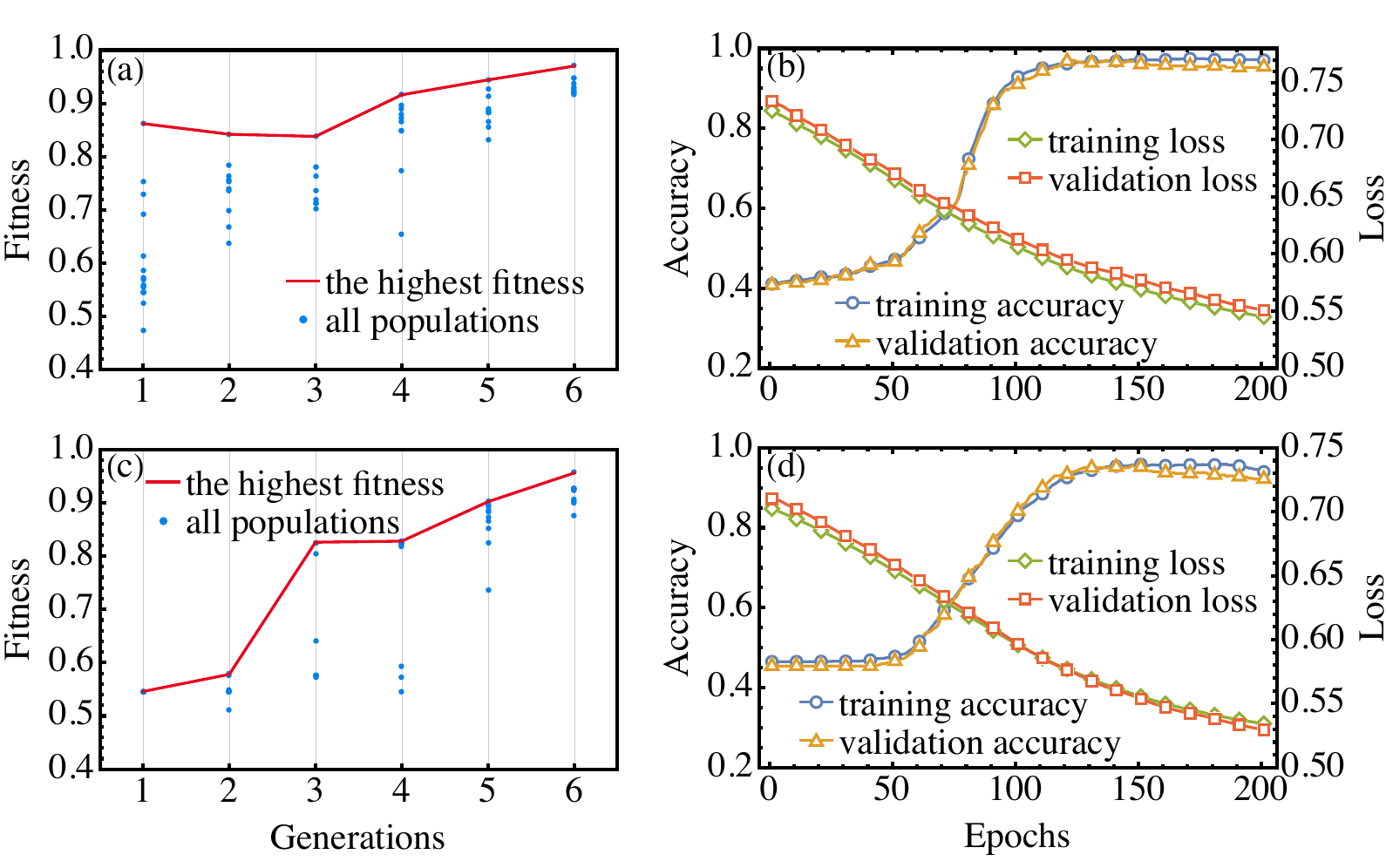}
\caption{Numerical results for classification of handwritten-digit images in the MNIST dataset. In (a), the initial variational parameters of single-qubit rotation gates are randomly chosen during the training process. In (b), we plot the loss and accuracy as a function of training epochs for the sixth-generation quantum circuit with the highest fitness. (c) shows the fitness of generated circuits for each generation with fixed initial variational parameters and (d) plots the loss and accuracy  for the sixth-generation circuit that has largest fitness \cite{Note1}. 
}
\label{fig:MNIST}
\end{figure}

\section{Classification of handwritten-digit images}

The first example we consider is the classification of handwritten-digit images in the MNIST dataset. This is a prototypical machine-learning task for benchmarking the effectiveness of various learning approaches. 
The MNIST dataset consists of gray-scale images for handwritten digits from $0$ through $9$. Each image is two dimensional, and contains $28\times 28$ pixels. In the Supplemental Material \cite{Note1}, we display four gray-scale images for digits 1, 4, 7 and 9 from the MNIST dataset.
For our purpose, we choose only a subset of MNIST consisting of images for digits $1$ and $9$ and reduce the size of the images to $16 \times 16$ pixels, so that we can run our MQNE algorithm and simulate the quantum classifiers generated with moderate classical computational resources. We use amplitude encoding  to map the input images into quantum states and define the following loss function based on cross entropy for a single data sample encoded as $|\psi\rangle_{\text{in}}$ (see \cite{Lu2020Quantum} for more details):
\begin{eqnarray}
L\left(h\left(|\psi\rangle_{\mathrm{in}} ; \Theta\right), \mathbf{a}\right)=-a_1\log g_1-a_2\log g_2, \label{Eq:loss}
\end{eqnarray}
where $\mathbf{a} = (a_1, a_2)=(1, 0) \text{ or } (0, 1)$ denotes the one-hot encoding \cite{goodfellow2016deep} of the label of $|\psi\rangle_{\text{in}}$, $h\left(|\psi\rangle_{\mathrm{in}} ; \Theta\right)$ represents the output of the quantum classifier with its parameters denoted by $\Theta$ collectively, and $g_{1,2}$ denote the output probabilities of digits $1$ and $9$.  For training the quantum classifier, we use a classical optimizer to search the optimal parameters $\Theta^*$ that minimize the averaged loss over the training dataset. 

For images with $16 \times 16$ pixels, we need eight qubits to encode each input sample, and for convenience, we also use an additional qubit to output the results of the binary classification. Thus, the ansatz circuit we aim to design is a nine-qubit variation circuit.   Applying the graph-encoding method and supposing that controlled-$R_x$ gates act only on adjacent qubits, we obtain $6688$ gate blocks and the corresponding directed graph has  $6688$ nodes. Based on the connection rules, we compute the adjacency matrix and apply the MQNE algorithm with hyperparameters set as $(n_i,t_i,l,l')=(10,1,5,2)$. Our results are summarized in Fig.~\ref{fig:MNIST}. In Fig.~\ref{fig:MNIST}(a), we randomly choose the initial variational parameters for single-qubit
rotation gates when training the generated quantum classifiers at each generation. The MQNE algorithm outputs a quantum circuit with fitness (accuracy) $97\%$ at the sixth generation, whose circuit structure is explicitly shown within the Supplemental Material \cite{Note1}.  The corresponding path for this circuit on  the directed graph reads 
$ [6687 \rightarrow 3969 \rightarrow 4418 \rightarrow 1321  \rightarrow 6319  \rightarrow 2817  \rightarrow 5933  \rightarrow 859  \rightarrow 2183  \rightarrow 5160  \rightarrow 4641  \rightarrow 4097  \rightarrow 4388  \rightarrow 2305  \rightarrow 4388] $,
where the numbers denote the labels of the nodes of the graph. In Fig.~\ref{fig:MNIST}(b), we plot the average accuracy and loss for both the training and validation datasets as a function of epochs during the training process. After training, the performance of this quantum classifier is also tested on the testing dataset and an accuracy of $97\%$ is obtained. Fig.\ref{fig:MNIST} (c) and (d) are analogous to  Fig.\ref{fig:MNIST} (a) and (b), respectively, but with fixed initial parameters for single-qubit rotation gates during the training process. We find that fixing the initial parameters would lead to a more stable improvement of the performance for next-generation classifiers.

\begin{figure}
\includegraphics[width = 0.45\textwidth]{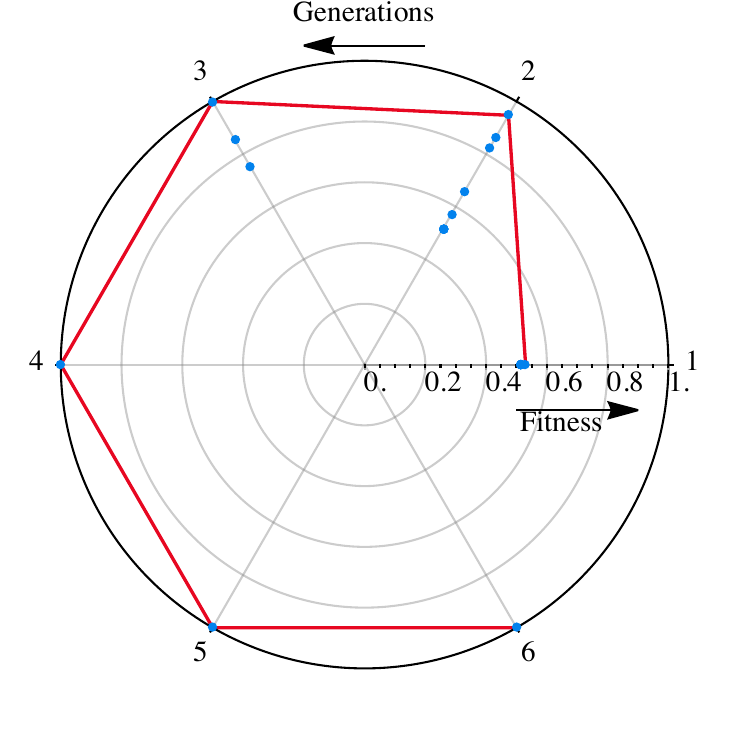}
\caption{Performance of the MQNE algorithm in the task of classifying symmetry-protected topological states \cite{Note1}. 
}  
\label{fig:SPT}
\end{figure}

\section{Classification of symmetry-protected topological states}

Unlike classical classifiers, quantum classifiers may also be used to directly classify quantum states  produced by quantum devices. To show the power of our MQNE algorithm in this scenario, we consider a quantum machine-learning task of classifying SPT states. For simplicity and concreteness, we consider the following cluster-Ising model, whose Hamiltonian reads \cite{Smacchia2011Statistical}
\begin{eqnarray}
H(\lambda)=-\sum_{j=1}^{N} \sigma_{j-1}^{x} \sigma_{j}^{z} \sigma_{j+1}^{x}+\lambda \sum_{j=1}^{N} \sigma_{j}^{y} \sigma_{j+1}^{y},
\end{eqnarray}
where $\sigma_i^{\alpha}$, $\alpha = x, y, z,$ are Pauli matrices acting on the $i$-th spin and $\lambda$ is a parameter describing the strength of the nearest-neighbour interaction, and $N$ denotes the number of spins. This model is exactly solvable and features a well-understood quantum phase transition at $\lambda = 1$, between a $Z_2\times Z_2$ SPT cluster phase characterized by a string order  for $\lambda < 1$ and an antiferromagnetic phase with long-range order for $\lambda > 1$. Here, we apply the MQNE algorithm to obtain an optimal ansatz variational circuit, which serves as a quantum classifier for classifying these two distinct phases. To this end, we set $N=8$ and uniformly sample $2000$ Hamiltonians with varying $\lambda$ from $0$ to $2$ under the periodic boundary condition. We compute their corresponding ground states, which are the input data to the classifier, and randomly choose $1600$ of them for training and the remaining ones for testing. Our results are plotted in Fig.~\ref{fig:SPT}, from which it is evident that the largest fitness increases at the first several generations and then saturates. We find a circuit at the third generation,  which involves only $36$ single-qubit and $10$ two-qubit gates but has a fitness equals $100\%$ \cite{Note1}.

We stress that, in comparison with the typical variational circuits used in previous works  \cite{Lu2020Quantum},  the ansatz circuits found by the MQNE algorithm involve much fewer gates and variational parameters, while maintaining a comparable classification accuracy.  
For instance, for the example of classification of handwritten-digit images, the classifier used in Ref.\cite{Lu2020Quantum} uses more than $90$ single-qubit and $80$ two-qubit  gates with circuit depth larger than $30$ and the number of variational parameters larger than $270$, whereas the circuit found by the MQNE algorithm at the sixth generation  contains only $28$ single-qubit rotation gates and $22$ two-qubit gates with $106$ variational parameters and circuit depth $15$.

This significant reduction of the circuit depth and number of gates (up to a constant factor) is crucial for experimental demonstration of quantum learning with NISQ devices, as the depth of the quantum circuits would be limited due to undesirable noises carried by such devices.
It not only simplifies the implementation of quantum classifiers substantially from the practical perspective, but also would mitigate the possible barren plateau problem (i.e., vanishing gradient) \cite{McClean2018Barren,Cerezo2020CostFunctionDependent,Grant2019Initialization} in training deep networks. We also mention that the performance of the MQNE may be improved further by choosing the hyperparameters judiciously according to different learning problems and  experimental setups. 
In the Supplemental Material, we also tested the MQNE algorithm in the task of classification of images from the Wisconsin Diagnostic Breast Cancer dataset, which may have useful application in medical machine learning \cite{Erickson2017Machine}.

\section{Discussion and conclusion}
Recent advances in quantum machine learning have revealed that quantum classifiers are highly vulnerable to adversarial attacks---adding a tiny amount of carefully crafted perturbations into the original legitimate data will cause the quantum classifiers to make incorrect predictions \cite{Lu2020Quantum,Liu2020Vulnerability}.  Thus, how to enhance the robustness of quantum classifiers to adversarial perturbations is a problem for practical applications of quantum learning in the future. With the MQNE algorithm,  a possible solution to this problem is to design ansatz circuits that are more robust to the given type of adversarial attack. This could be achieved by replacing the original loss function [e.g., Eq. (\ref{Eq:loss})] with a modified one that incorporates the adversarial perturbations \cite{Chakraborty2018Adversarial}. In addition, the graph-encoding method would also be combined with other evolution or genetic algorithms to construct optimal circuit structures for different quantum-learning problems. 
In the future, it would be interesting to consider some symmetries in the data as prior knowledge to enhance our algorithm by restricting the searching space into a smaller subspace.

In summary, we introduce a quantum neuroevolution algorithm, named the MQNE algorithm, to design optimal variational  ansatz quantum circuits for different quantum-learning tasks.
Through concrete examples involving classifications of real-life images and SPT quantum states, we demonstrate that the MQNE algorithm performs excellently in searching appropriate quantum classifiers. It finds ansatz circuits with notably smaller depths and number of gates, while maintaining a comparable classification accuracy. Our results provide  a valuable guide for experimental implementations of quantum machine learning with NISQ devices. 

We acknowledge helpful discussions with  Weikang Li, Wenjie Jiang, and Sirui Lu. This work is supported by the start-up fund from Tsinghua University (Grant. No. 53330300320), the National Natural Science Foundation of China (Grant. No. 12075128), and the Shanghai Qi Zhi Institute. 

\bibliography{Q-neuroevolution}

\clearpage

\setcounter{secnumdepth}{3}

\makeatletter
\renewcommand{\thefigure}{S\@arabic\c@figure}
\renewcommand \theequation{S\@arabic\c@equation}
\renewcommand \thetable{S\@arabic\c@table}
\renewcommand \thealgorithm{S\@arabic\c@algorithm}

\begin{center} 
	{\large \bf Supplemental Material: Markovian Quantum Neuroevolution for Machine Learning}
\end{center}

In this Supplemental Material, we specifically show how to narrow (enlarge) the gate-block library by imposing (cancelling) some restrictions on building possible gate-blocks.
We also mention how to extend the Markovian process in the MQNE algorithm to the high-order Markovian process.
Besides, we present more details on the graph-encoding method, the MQNE algorithm, and more numerical results to demonstrate the performance of the proposed scheme.

\section{Encoding Vectors and The Pseudocode}

In the main text, we restrict our discussion to the case that the circuits are composed with only single-qubit rotation gates $R$ (composed of $R_z R_x R_z$, where $R_x$ and $R_z$ denote rotations along $x$ and $z$ axes, respectively) and two-qubit controlled-$R_x$ gates,  and the controlled-$R_x$ gates only act on adjacent qubits.  The pseudocode for the MQNE algorithm is provided in Algorithm \ref{alg:MQNE algorithm}.

\begin{figure}
    \begin{algorithm}[H]
    \caption{Markovian quantum neuroevolution algorithm}
    \label{alg:MQNE algorithm}
    \begin{algorithmic}
    \REQUIRE The directed graph, hyperparameters $n_i$, $t_i$, $l$, $l'$, $f_c$, and $g_c$
    \ENSURE The optimal quantum circuit architecture
    \STATE Initialization: start from a fixed node and uniformly sample $n_1$ paths with length-$l$, and compute their fitness
    \FOR{ $i=1$ to $g_c$}
     \STATE Choose the $t_i$ ($t_i < n_i$) paths with highest fitness
     \STATE Evolution: produce paths in the $(i+1)$-th generation by concatenating  $n_{i+1}$ segments of length $l'$ to each of the $t_i$ paths, and compute their fitness 
     \IF{ $\max$ [fitness (paths)] $\geq f_c$ }
             \STATE Terminate the iteration
     \ENDIF
    \ENDFOR
    \STATE Output the optimal quantum circuit
    \end{algorithmic}
    \end{algorithm}
\end{figure}

\section{The gate-block library}

\begin{figure}[t]
    \includegraphics[width = 0.48\textwidth]{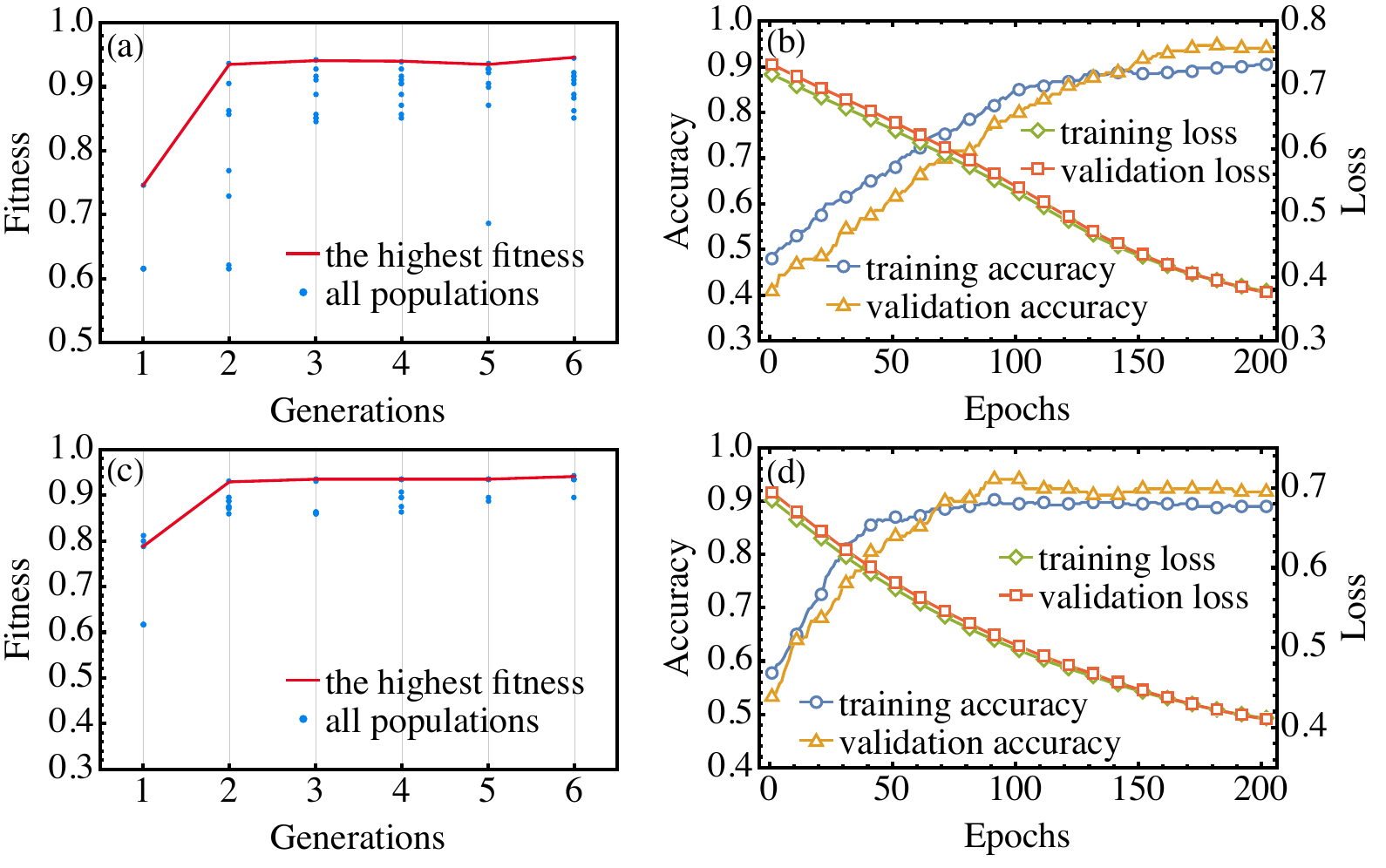}
    \caption{The numerical results for applying the MQNE algorithm to the task of classifying samples in the cancer dataset. In (a), we randomly choose the initial variational parameters of single-qubit rotation gates at the beginning of the training process. In (b), the loss and accuracy are plotted as functions of training epochs for the sixth-generation quantum circuit with the highest fitness. (c) and (d) are similar to (a) and (b), but with fixed initial variation parameters.  
    }  
    \label{fig:cancer dataset}
\end{figure}

For a $k$-qubit quantum gate-block library, we can sort all quantum gate-blocks into the ones containing $0$ controlled-$R_x$ gate, $1$ controlled-$R_x$ gate, until $\lfloor k/2 \rfloor$ controlled-$R_x$ gates.

For quantum gate-blocks containing $0$ controlled-$R_x$ gate, each of $k$ qubits can be acted on by a single-qubit rotation gate $R$ or an identity gate, corresponding to $2^{k}$ quantum gate-blocks of this type totally.
Likewise, for quantum gate-blocks containing $1$ controlled-$R_x$ gate, each of the remaining $k-2$ qubits can be acted on by rotation gate $R$ or identity gate pertaining to $2^{k-2}$ possibilities. Through some combinatorial calculations, we can obtain the total number of $1$ controlled-$R_x$ blocks:
$$
N_1(1) = 2^{k-2}\times\tbinom{k-1}{1}\times2^{1} \, .
$$
Similarly, for quantum gate-blocks containing $i$ controlled-$R_x$ gate, where $i\leq \lfloor k/2 \rfloor$, each of the remaining $k-2i$ qubits can be acted on by gate $R$ or the identity gate, we obtain
$$
N_1(i) = 2^{k-2i}\times\tbinom{k-i}{i}\times2^{i} \, .
$$
Then the total number of quantum gate-blocks is:
$$
f_1(k) = \sum_{i=0}^{\lfloor k/2 \rfloor}N_1(i) = \frac{(1+\sqrt{3})^{k+1} - (1-\sqrt{3})^{k+1}}{2\sqrt{3}} \ .
$$

\begin{figure}[t]
\includegraphics[width = 0.45\textwidth]{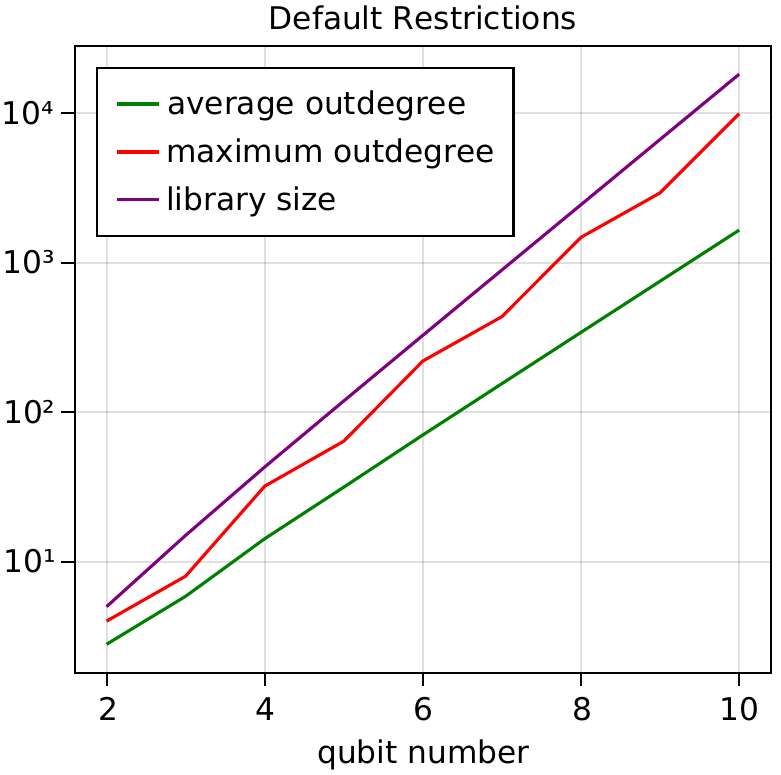}
\caption{ The size of the gate-block library, and the maximum (average) outdegree of all nodes in the directed graph as functions of qubit number under default restrictions in the main text.} 
\label{fig:exp}
\end{figure}

In this way, we have calculated there are  $f_1(k) = \Theta[(1+\sqrt{3})^k]$ quantum gate-blocks in the gate-block library used in the main text. To analyse the complexity of the searching space in the MQNE algorithm, we also numerically study how the outdegree (for a node, its outdegree is the number of other nodes it is allowed to point to) of nodes scale with respect to qubit number and show numerical results in Fig.~\ref{fig:exp}, from which it is evident that the maximum (average) value of the outdegree has the same growing trend as the library size, which will climb exponentially as qubit number increases.

In some realistic problems, considering limited classical computational power or other realistic reasons, especially for large qubit number case, it is necessary to impose further restrictions to reduce the size of the library, from the exponential scaling $f_1(k)$ to a polynomial scaling $f_2(k)$. in our proposal, We plus three more restrictions. Firstly, for quantum gate-blocks containing $i$ controlled-$R_x$ gate, we assume the remaining $k-2i$ qubits are acted on \textit{simultaneously} by either gate $R$ or the identity gates. Secondly, for each controlled-$R_x$ gate acting on two neighbouring qubits $i$ and $i+1$, we assume that the $i$-th qubit is the controlling qubit. Thirdly, we restrict each quantum gate-block contains at most $c$ (a cut-off constant) controlled-$R_x$ gates.

With these further restrictions, there are $N_2(i) = \tbinom{k-i}{i} $ gate-blocks containing $i$ controlled-$R_x$ gates, and consequently the total number of  gate-blocks $f_2(k) $ in the library is given by 
\begin{eqnarray*}
f_2(k) = \sum_{i=0}^{c}N_2(i) = \sum_{i=0}^{c}\tbinom{k-i}{i} = O(k^c)  .
\end{eqnarray*}

\begin{figure}[t]
\includegraphics[width = 0.45\textwidth]{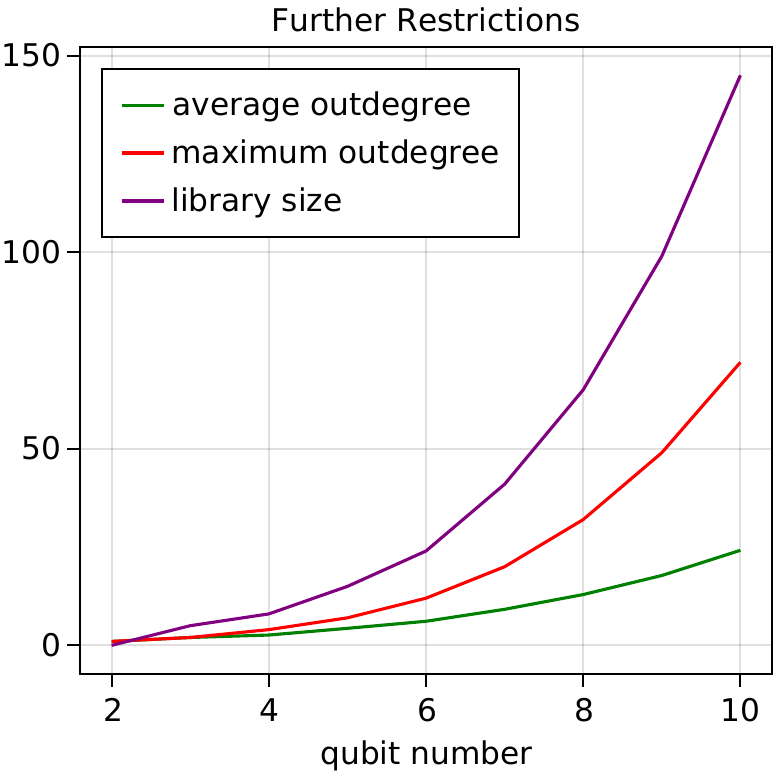}
\caption{ We plot the size of the gate-block library, and the maximum (average) outdegree of all nodes in the directed graph as functions of qubit number under further restrictions we added.} 
\label{fig:poly}
\end{figure}

And we also numerically explore how the outdegree scale with respect to qubit number with numerical results shown in Fig.~\ref{fig:poly}, which reflects that the maximum (average) value of the outdegree which will rise polynomially with respect to qubit number.

\begin{figure}[t]
\includegraphics[width = 0.48\textwidth]{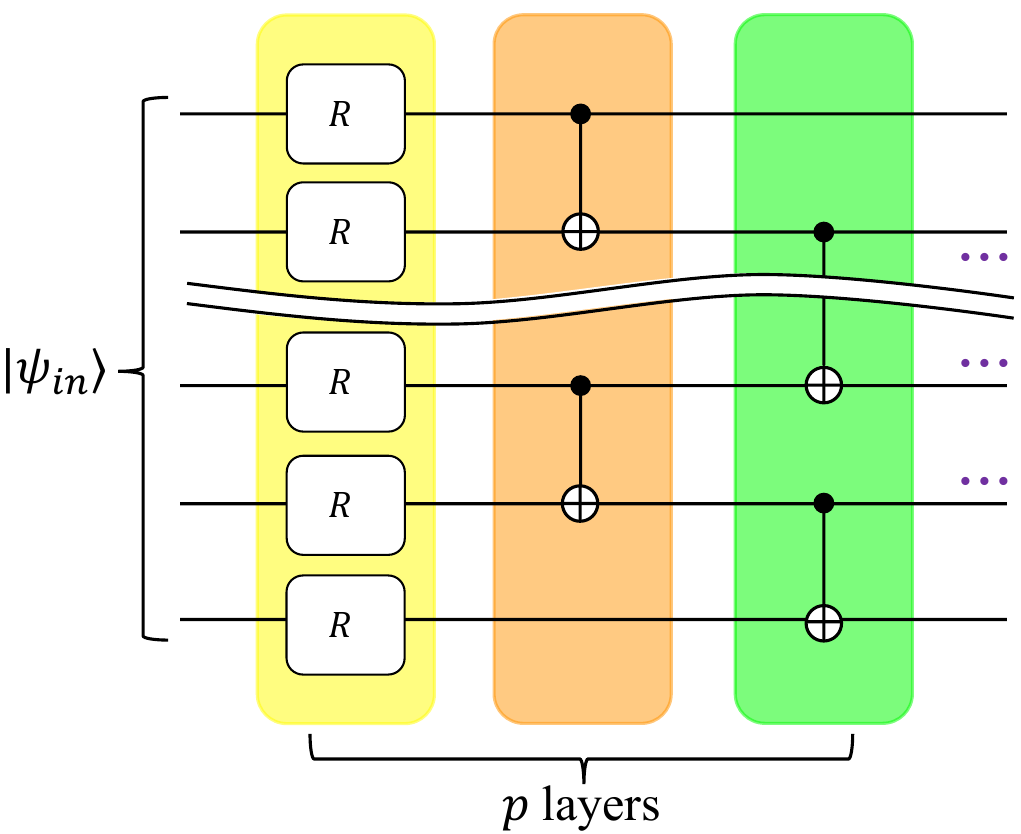}
\caption{The sketch of a quantum classifier architecture. It consists of p same layers, each layer has three gate-blocks, the gate-block marked yellow is the rotation unit, the gate-blocks marked orange and green are in the entanglement unit.} 
\label{fig:classifier}
\end{figure}

Similarly, adding more restrictions can even reduce the size of the library to a constant $f_3$, independent of $k$. 
For instance, a gate-block library containing only three gate-blocks can be constructed to obtained a quantum classifier architecture as shown in Fig.~\ref{fig:classifier}. 

We mention that, the gate-block marked yellow corresponds to the rotation layer, the gate-blocks marked orange and green correspond to the entanglement layer respectively, in the Fig.2 of Ref.~\cite{Lu2020Quantum}. 

On the contrary, in order to improve the possibility of finding the optimal quantum circuits, we can enlarge the searching space by releasing  some restrictions mentioned in the main text and above.
For example, we may remove the restriction that the controlled-$R_x$ gates only act on adjacent qubits. We  obtain the total number of quantum gate-blocks containing $1$ controlled-$R_x$ gate:
$$
N_0(1) = 2^{k-2}\times\mathrm{A}_k^2 \, .
$$
Likewise, for quantum gate-blocks containing $i$ controlled-$R_x$ gates, where $i\leq \lfloor k/2 \rfloor$, with each of the remaining $k-2i$ qubits being acted on by gate $R$ or identity gate, we obtain
$$
N_0(i) = 2^{k-2i}\times\mathrm{A}_k^{2i}/\mathrm{A}_i^i \, .
$$
Then the total number of gate-blocks is
$$
f_0(k) = \sum_{i=0}^{\lfloor k/2 \rfloor}N_0(i)  \,  .
$$
In this way, the size of the quantum gate-block library is enlarged to $\Omega(\lfloor k/2 \rfloor!)$. 

In practical applications, we can directly construct the gate-block library for $(k+1)$-qubit circuits based on existed libraries for $(k-1)$- and $k$-qubit circuits. Firstly, the operations on the top $k$ qubits can be adopted entirely from the gate-block library of $k$-qubit circuits when the last qubit is acted on by gate $R$ or identity gate.
Secondly, 
the operations on the top $(k-1)$ qubits can be adopted entirely from the gate-block library of $(k-1)$-qubit circuits when the last two qubits is acted on by a controlled-$R_x$ gate.
In this way, we can readily scale up the gate-block library of $(k-1)$-qubit and $k$-qubit circuits to the gate-block library of $(k+1)$-qubit circuits.

\section{The MQNE algorithm with high-order Markovian process}

In the main text, we consider the rules between adjacent gate-blocks with the help of the single-qubit rotation gate $R$. 
We could regard finding paths in the directed graph as a Markovian process, in which the future state depends only on the current state of the system, but not on the  previous ones. 
All paths of the graph correspond to the discrete state space, and the adjacency matrix of the graph corresponds to the transition matrix describing the probabilities of transitions between states.
In the main text, we consider a simple case where the transition probability is uniformly distributed on all states that the present state can transit to. 

For the high-order Markovian process, we could directly use $R_z$  and $R_x$ gates, instead of the composed $R$ gate, to construct quantum gate-blocks. Consequently, we need to invoke new rules between adjacent four gate-blocks to avoid ambiguity and duplication of successive rotations.
For example, when three consecutive quantum gate-blocks contain $R_z$, $R_x$, $R_z$ on the same qubit respectively, then the fourth gate-block should not have $R_z$ or $R_x$ on this qubit.
So the future state depends on the previous three states of the system, which is a third-order Markovian process.

\begin{figure}[t]
\includegraphics[width = 0.45\textwidth]{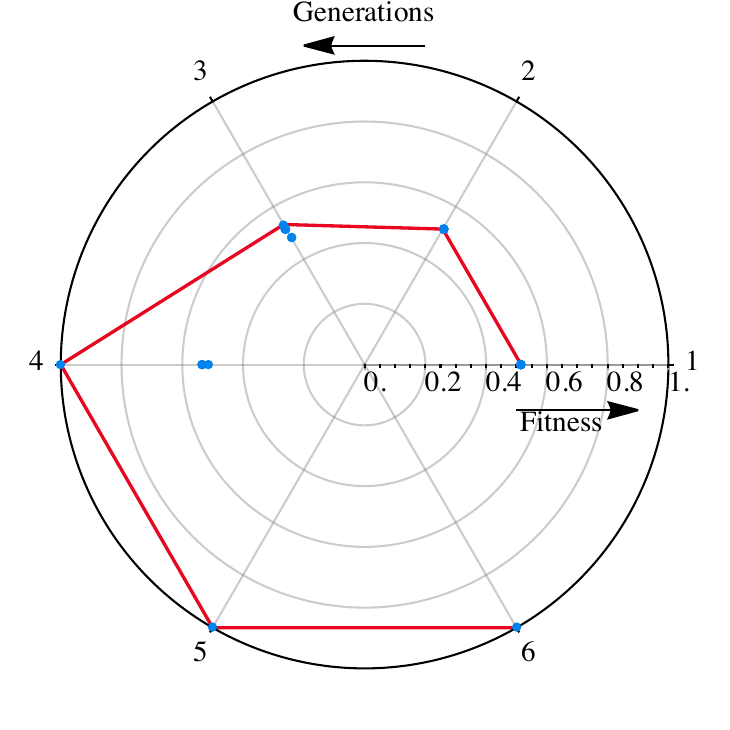}
\caption{The performance of the MQNE algorithm in the task of classifying symmetry-protected topological states, with the initial variational parameters fixed at the beginning of the training process.
}  
\label{fig:SPT2-Evolution}
\end{figure}

\section{numerical results for the cancer dataset}
Another example we consider is the classification of samples in the cancer dataset \cite{cancer}, we need six qubits to encode each input sample and use an additional qubit to output the results of the binary classification (whether the label is cancer or not).
Thus, the ansatz circuit we aim to design is a seven-qubit variational circuit. Applying the graph-encoding method for seven-qubit circuits and supposing that controlled-$R_x$ gates act only on adjacent qubits, we obtain $896$ gate-blocks and the corresponding directed graph has $896$ nodes. Based on the connection rules, we compute the adjacency matrix and apply the MQNE algorithm with hyperparameters set as $(n_i,t_i,l,l')=(10,1,3,2)$.

Our results are summarized in Fig.~\ref{fig:cancer dataset}. In Fig.~\ref{fig:cancer dataset}(a), we randomly choose the initial variational parameters for single-qubit rotation gates when training the generated quantum classifiers at each generation. The MQNE algorithm outputs a quantum circuit (with $68$ variational parameters) with fitness  $94.6\%$ at the sixth generation. The corresponding path for this circuit on the directed graph reads $ [895 \rightarrow 31 \rightarrow 801 \rightarrow 273 \rightarrow 404 \rightarrow 643 \rightarrow 423 \rightarrow 593 \rightarrow 225 \rightarrow 481 \rightarrow 673 \rightarrow 792 \rightarrow 609] $, where the numbers denotes the labels of the nodes of the graph.

In Fig.~\ref{fig:cancer dataset}(b), we plot the average accuracy and loss for both the training and validation datasets as a function of the number of epochs during the training process. After training, the performance of this quantum classifier is also tested on the testing dataset and an accuracy of $94.6\%$ is obtained. We mention that the numerical simulations of training the quantum classifiers are based on  the Julia language \cite{Bezanson2017Julia} and Yao.jl \cite{Luo2020Yao} framework, throughout the paper.
Fig.~\ref{fig:cancer dataset} (c) and (d) are analogous to  Fig.~\ref{fig:cancer dataset} (a) and (b) respectively, but with fixed initial parameters during the training process.

\begin{figure}[t]
\includegraphics[width = 0.48\textwidth]{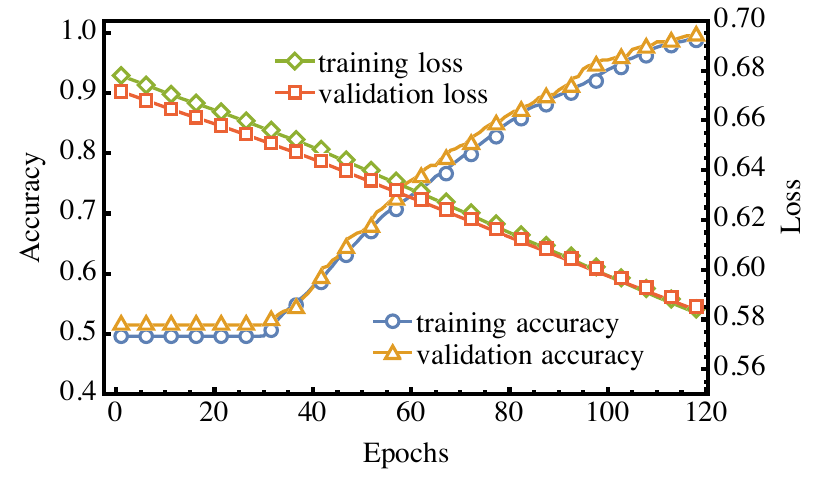}
\caption{The numerical results for classifying symmetry-protected topological states
of the cluster-Ising model. 
The initial variational parameters of single-qubit rotation gates are randomly chosen during the training process, we plot the loss and accuracy as a function of training epochs for the sixth-generation quantum circuit with the highest fitness. 
}  
\label{fig:SPT1-Training}
\end{figure}

\section{more numerical results for classifying handwritten-digits and SPT states}

The MNIST dataset consists of two-dimensional gray-scale images ($28\times 28$ pixels) for handwritten digits from $0$ through $9$, here we display four gray-scale images for digits 1, 4, 7 and 9 from MNIST dataset in Fig.~\ref{fig:digits}. 

In the classification of handwritten-digit images in the MNIST dataset, the MQNE algorithm outputs a quantum circuit with fitness (accuracy) 97\% at the sixth generation, whose structure is explicitly shown in Fig.~\ref{fig:circuit-MNIST}.

In the main text, we also consider a quantum machine learning task of classifying SPT states. We use two strategies to train the generated quantum classifiers at each generation.
Firstly, we randomly choose the initial variational parameters for single-qubit rotation gates when training the generated quantum classifiers at each generation, our results are plotted in the main text. The MQNE algorithm outputs a quantum circuit (with $46$ variational parameters) with fitness   $100\%$ at the third generation. The corresponding path for this circuit on the directed graph reads $ [6687 \rightarrow 1665 \rightarrow 2177 \rightarrow 2401 \rightarrow 4899 \rightarrow 3042 \rightarrow 4901] $.
In Fig.~\ref{fig:SPT1-Training}, we plot the average accuracy and loss for both the training and validation datasets as a function of the number of epochs during the training process.

\begin{figure}[t]
\includegraphics[width = 0.48\textwidth]{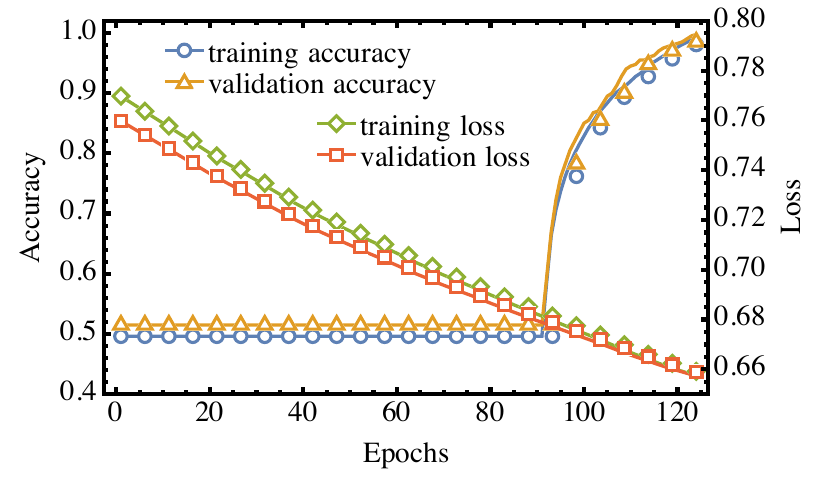}
\caption{The loss and accuracy as a function of training epochs for the sixth-generation circuit with the highest fitness (see Fig.~\ref{fig:SPT2-Evolution}), for the task of classifying symmetry-protected topological states of the cluster-Ising model.}  
\label{fig:SPT2-Training}
\end{figure}

Secondly, we fix initial parameters for single-qubit rotation gates during the training process, the results are shown in  Fig.~\ref{fig:SPT2-Evolution}. The MQNE algorithm outputs a quantum circuit (bearing $54$ variational parameters) with fitness  $100\%$ at the fourth generation. The corresponding path for this circuit on the directed graph reads $ [6687 \rightarrow 3649 \rightarrow 3173 \rightarrow 737 \rightarrow 1798 \rightarrow 60 \rightarrow 1217 \rightarrow 4905 \rightarrow 1826]$, 
In Fig.~\ref{fig:SPT2-Training}, we plot the average accuracy and loss for both the training and validation datasets as a function of the number of epochs during the training process. 

\section{parameter settings in training quantum classifiers}

In the task for classification of handwritten-digit images in the MNIST dataset (see Fig.~2 in the main text),
we use the Adam optimizer with a batch size of $30$ and a learning rate of $0.0015$ to minimize the loss function. The training step is $200$. The fitness is averaged on $2000$ training samples and $500$ validation samples which are not contained in the training dataset. 

In the task for classification of samples in the cancer dataset (see Fig.~\ref{fig:cancer dataset}),  
we use the Adam optimizer with a batch size of $20$ and a learning rate of $0.0015$ to minimize the loss function. The training step is $150$. The fitness is averaged on $400$ training samples and $169$ validation samples which are not contained in the training dataset.  

In the task for classifying symmetry-protected topological states (see Fig.~3 in the main text, and Figs. \ref{fig:SPT2-Evolution}, \ref{fig:SPT2-Training} and \ref{fig:SPT1-Training} in this Supplemental Material),
we use the Adam optimizer with a batch size of $20$ and a learning rate of $0.0015$ to minimize the loss function. The training step is $150$. The fitness is averaged on $1600$ training samples and $400$ validation samples which are not contained in the training dataset.  

\section{The direct generalization of the NEAT algorithm in searching quantum circuits}

In this part, we show the performance of the genetic algorithm, which could be regarded as a naive generalization of the NEAT algorithm \cite{Stanley2002Evolving},  in searching quantum circuits.

In conventional evolutionary algorithms, we evolve the population to create the next generation by applying genetic operators on individuals to generate offspring.  Two important genetic operators are mutation and crossover \cite{Stanley2002Evolving, Stanley2019Designing}, where crossover produce new offspring by recombining two selected individuals (parents), and mutation generate new offspring by randomly mutating a selected individual. Applying them in searching quantum circuits, crossover is defined to divide each of two parent quantum circuits into two parts respectively and exchange their divided parts to create two offspring, mutation is defined to randomly generate some positions in circuits and then replace quantum gates in these positions with other different gates.
\begin{figure}[t]
\includegraphics[width = 0.45\textwidth]{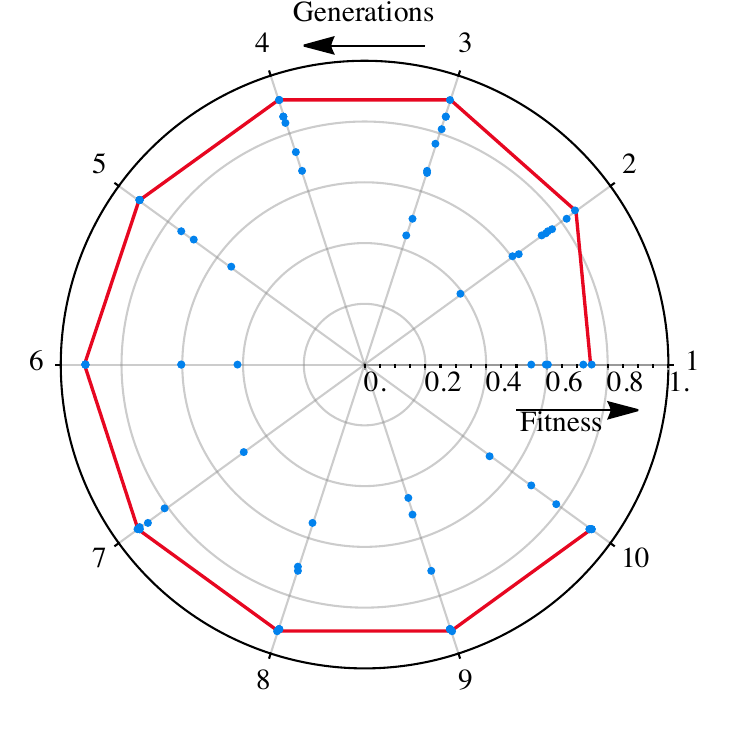}
\caption{ The performance of the genetic algorithm in the task of classifying handwritten-digit images in the MNIST dataset.}  
\label{fig:operator_pop}
\end{figure}

\begin{figure}[t]
\includegraphics[width = 0.48\textwidth]{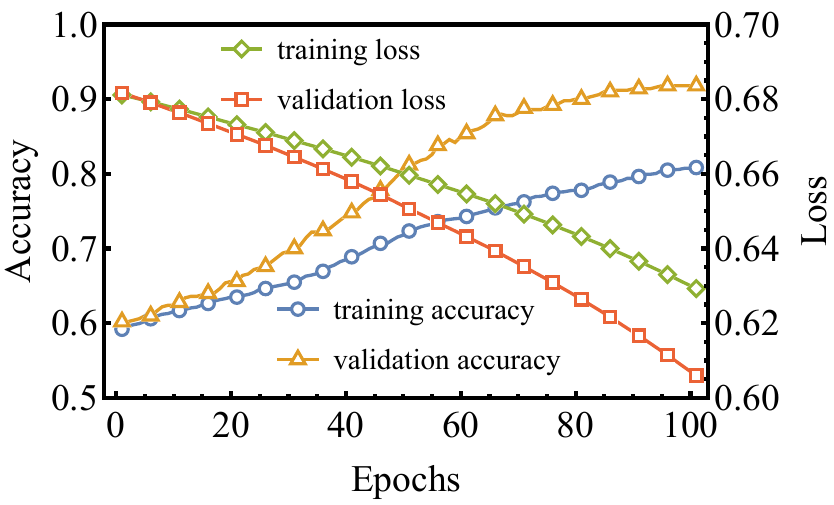}
\caption{The loss and accuracy as a function of training epochs for the tenth-generation quantum classifier with the highest fitness, in classifying handwritten-digit images in the MNIST dataset. Here, we use the genetic algorithm to construct quantum circuits. }  
\label{fig:operator_train}
\end{figure}

\begin{figure}[h]
    \includegraphics[width = 0.48\textwidth]{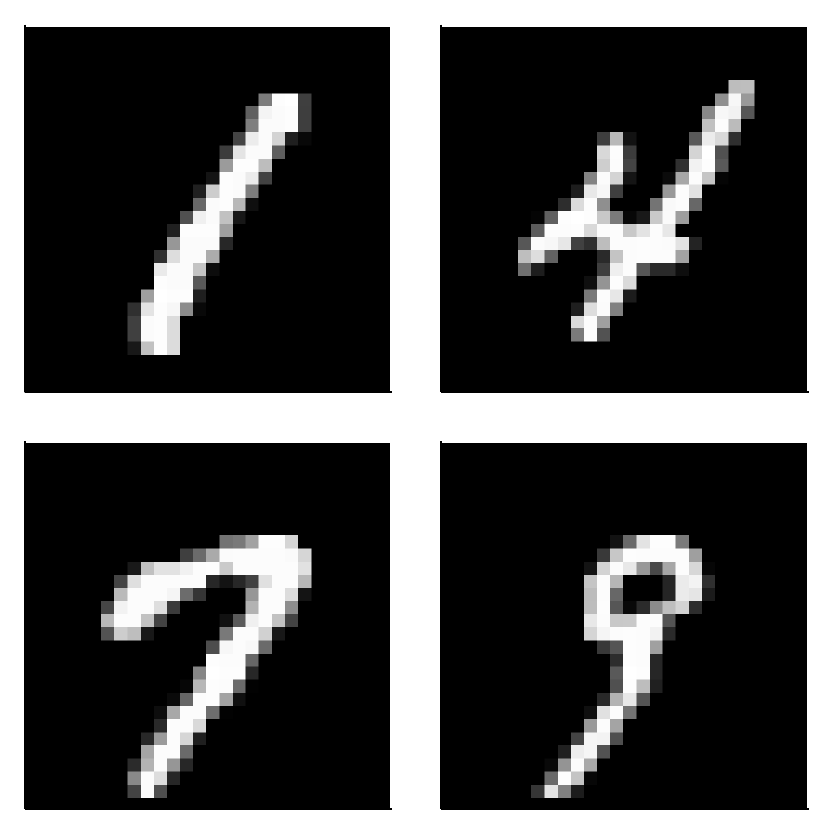}
    \caption{ Images for handwritten digits from MNIST dataset: 1, 4, 7, 9 \cite{mnist}.}  
    \label{fig:digits}
\end{figure}

We design the genetic algorithm as shown in Algorithm~\ref{alg:evolution algorithm} to search for optimal quantum classifier architectures (9-qubit circuits) for the MNIST handwritten digit  dataset. The major hyperparameters   ($n_i$, $t_i$) are set as ($9$, $3$).

Fig.~\ref{fig:operator_pop} displays the fitness of $86$ quantum circuits totally evaluated in running the genetic algorithm. We find a circuit structure with fitness $92\%$ at the third generation, whose number of quantum gates is more than $100$.
In Fig.~\ref{fig:operator_train}, we plot the average accuracy and loss for both the training and validation datasets as a function of the number of epochs during the training process. 
We see from Fig.~\ref{fig:operator_pop} that the local convergence appears at the third generation and the fitness does not increase for later generations. The performance of the genetic algorithm is ineffective when directly applying these two operators in designing quantum classifiers.

\begin{figure}[b]
    \begin{algorithm}[H]
    \caption{The genetic algorithm}
    \label{alg:evolution algorithm}
    \begin{algorithmic}
    \REQUIRE Hyperparameters $n_i$, $t_i$, $g_c$, $f_c$, etc
    \ENSURE The optimal quantum circuit architecture
    \STATE Initialization:: randomly generate $n_1$ quantum circuits, and compute their fitness
    \FOR{ $i=1$ to $g_c$}
    \STATE Choose the best $t_i$ ($t_i < n_i$) circuits with highest fitness
    \STATE Apply the crossover and mutation operator
    \STATE Compute the fitness of individuals in the $(i+1)$-th generation
    \IF{ max[fitness(circuits)] $\geq f_c$}
            \STATE Terminate the iteration
    \ENDIF
    \ENDFOR
    \end{algorithmic}
    \end{algorithm}
\end{figure}

\begin{figure*}[t]
    \includegraphics[width = 0.96\textwidth]{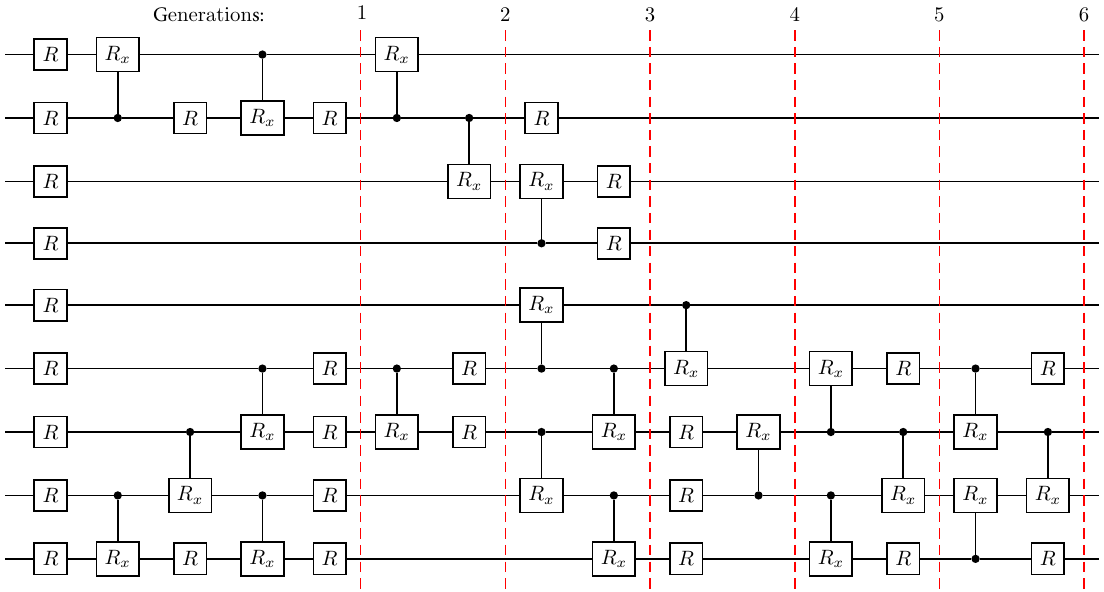}
    \caption{The optimal structure of the quantum circuit generated by the MQNE algorithm to classify handwritten-digit images in the MNIST dataset. The numbers in the top denote the generations.}  
    \label{fig:circuit-MNIST}
\end{figure*}

\newpage


\end{document}